\documentclass[prl,twocolumn]{revtex4}
\usepackage{graphicx}% Include figure files
\usepackage{dcolumn}% Align table columns on decimal point
\usepackage{amsmath,amsbsy,amssymb,graphics}
\usepackage{bm}% bold math
\usepackage{mathrsfs,color}

\newcommand{\imag}{\mathrm{Im}\,}

\newcommand{\imu}{\mathrm{i}}

\newcommand{\dg}{\dagger}

\newcommand{\sg}{\sigma}

\begin{document}

\preprint{APS/123-QED}

\title{
A spin-freezing perspective on cuprates
}

\author{Philipp Werner$^1$, Shintaro Hoshino$^{2}$ and Hiroshi Shinaoka$^{3}$}

\affiliation{
$^1$Department of Physics, University of Fribourg, 1700 Fribourg, Switzerland\\
$^2$ RIKEN Center for Emergent Matter Science (CEMS), Wako, Saitama 351-0198, Japan \\ 
$^3$ Department of Physics, Saitama University, Saitama 338-8570, Japan
}

\date{July 28, 2016}

\hyphenation{instabi-li-ty}

\begin{abstract}
The high-temperature superconducting state in cuprates appears if charge carriers are doped into a Mott insulating parent compound. An unresolved puzzle is the unconventional nature of the normal state above the superconducting dome, and its connection to the superconducting instability. At weak hole-doping, 
a ``pseudo-gap" metal state 
with signatures of time-reversal symmetry breaking 
is observed, which near optimal doping changes into a ``strange metal" with 
non-Fermi liquid properties. Qualitatively 
similar phase diagrams are found in multi-orbital systems, such as pnictides, where the unconventional metal states arise from a Hund coupling induced spin-freezing. Here, we show that the relevant model for cuprates, the single-orbital Hubbard model on the square lattice, can be mapped onto an effective multi-orbital problem with strong ferromagnetic Hund coupling.  
The spin-freezing physics of this multi-orbital system 
explains the phenomenology of cuprates, including the pseudo-gap, the strange metal, and the $d$-wave superconducting instability. Our analysis 
suggests  
that spin-freezing is the  
universal mechanism 
which controls the 
properties of unconventional superconductors.   
\end{abstract}

\pacs{Valid PACS appear here}
\maketitle

Hund coupling effects, in particular spin-freezing \cite{werner2008}, produce remarkable   
phenomena in correlated multi-orbital systems \cite{georges2013}. The emergence of magnetic moments in the correlated metal phase leads to characteristic non-Fermi-liquid properties \cite{werner2008}. At low enough temperature, the fluctuating local moments can trigger a symmetry-breaking to unconventional superconducting or excitonic states, which generically border a magnetically ordered phase \cite{hoshino2015,hoshino2016}. The spin-freezing crossover occurs in a narrowly defined range of fillings and interaction strengths, and the remarkable fact is that many unconventional multi-band superconductors fall into this parameter region. Examples are iron pnictides \cite{haule2009,liebsch2010,werner2012}, chromium based superconductors \cite{huang2016}, strontium ruthenates \cite{werner2008,georges2013,hoshino2016}, and uranium based compounds \cite{hoshino2015}. In fulleride superconductors \cite{capone2009}, where the effective Hund coupling is negative \cite{nomura2015} and the roles of spin and orbital are in some sense exchanged, the unconventional superconducting state is associated with an orbital-freezing phenomenon \cite{steiner2016}. A conspicuous exception from this almost exhaustive list of unconventional superconductors are the cuprates, which are typically 
discussed in terms of a single-band Hubbard model, where the Hund interaction does not appear. 

Here, we 
introduce a basis transformation which 
maps  
the two-dimensional (2D) single-orbital Hubbard model onto a   
two-orbital model with  
ferromagnetic Hund coupling. 
The spin-freezing physics of this two-orbital system explains the pseudo-gap and bad-metal state of the weakly doped Hubbard model, and the crossover to Fermi-liquid properties near optimal doping. We will also show that the slow local moment fluctuations 
associated with spin-freezing 
provide the glue for $d$-wave 
pairing.

\begin{figure*}[ht]
\begin{center}
\includegraphics[width=0.9\textwidth]{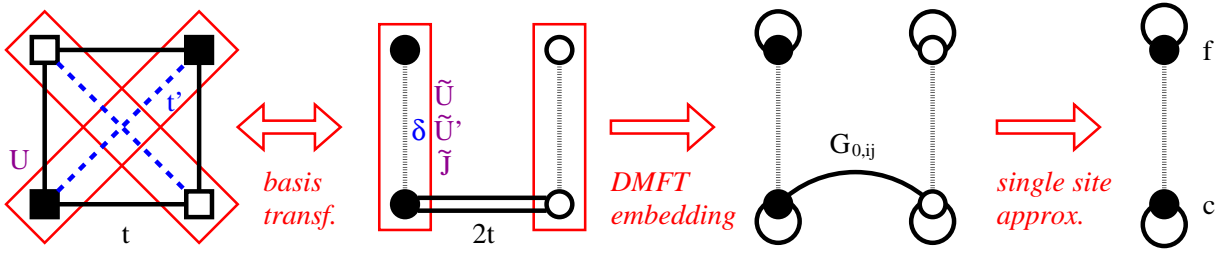}
\caption{
%(Color online)
Illustration of the mapping of the plaquette with nearest neighbor hopping $t$ and diagonal hopping $t'$ onto a coupled pair of two-orbital models. If $U$ is the on-site interaction on the plaquette, the two-orbital system has a Slater-Kanamori type interaction with $\tilde U=\tilde U'=\tilde J=U/2$. The diagonal hopping translates into a 
crystal-field splitting $\delta=2t'$. 
The third panel shows the self-consistent embedding of the two-orbital system into a noninteracting bath described by $(G_0)_{ij}$, and the right most panel the simplification to a single-site two-orbital impurity problem. 
}
\label{fig:illustration}
\end{center}
\end{figure*}

\vspace{2mm}\noindent
{\bf Model and Method}

We consider the 2D Hubbard model with on-site interaction $U$, nearest neighbor hopping $t$ and next-nearest neighbor hopping $t'$,
\begin{align}
\mathscr{H} &= \sum_i U n_{i\uparrow}n_{i\downarrow}  - \sum_i \mu (n_{i\uparrow}+n_{i\downarrow})\nonumber \\
&-t\sum_{\langle i,j\rangle,\sigma} (d^\dagger_{i\sigma} d_{j\sigma} + \text{h.c.})-t' \sum_{\langle\langle i,j\rangle\rangle,\sigma} (d^\dagger_{i\sigma} d_{j\sigma} + \text{h.c.}).
\label{hamilt}
\end{align}
Here, $i$ and $j$ are site indices, $\langle i,j\rangle$ denotes nearest neighbor pairs, and $\langle\langle i,j\rangle\rangle$ next-nearest neighbor pairs. The density operator is $n_\sigma=d^\dagger_\sigma d_\sigma$ and the chemical potential is $\mu$. This model is  
a fundamental model of cuprate superconductors, since it describes the physics of the copper-oxygen plane. More specifically, the single band corresponds to the strongly hybridized anti-bonding combination of Cu $d_{x^2-y^2}$ and O $p_x$ and $p_y$ orbitals. 
A typical parameter choice is 
$U\approx 8t$ and $t'\approx -0.3t$ \cite{pavarini2001}. Since there is only a single orbital per site, 
Hund coupling effects such as spin-freezing have not been discussed in connection with cuprates. 

Instead, because of the strong antiferromagnetic correlations and the $d$-wave nature of the superconducting state, the physics of the plaquette, illustrated in the left-hand panel of Fig.~\ref{hamilt}, plays a prominent role. This plaquette is the building block for 4-site cluster dynamical mean field theory (DMFT) calculations \cite{lichtenstein2000,kotliar2001}, which have been extensively used to investigate the 2D Hubbard model and which have produced phasediagrams in qualitative agreement with that of cuprates \cite{maier2005}. 

To analyze the physics of the plaquette and the 2D Hubbard model from a multi-orbital perspective, we perform a basis transformation to bonding/antibonding orbitals, as illustrated in Fig.~\ref{fig:illustration}. If the sites are numbered in an anti-clockwise fashion starting from the bottom left, the transformed orbitals are defined as follows:
\begin{align}
c_1&=\tfrac{1}{\sqrt{2}}(d_1+d_3), \quad
c_2=\tfrac{1}{\sqrt{2}}(d_2+d_4),\\
f_1&=\tfrac{1}{\sqrt{2}}(d_1-d_3), \quad
f_2=\tfrac{1}{\sqrt{2}}(d_2-d_4).
\end{align}
This transformation maps the plaquette onto a pair of two-orbital systems, with a hopping of $2t$ between the antibonding ($c$) orbitals, and no hopping between the bonding ($f$) orbitals. The interactions between the two orbitals on a given site are of the ``Slater-Kanamori" type, 
\begin{align}
\tilde{\mathscr{H}} _\text{loc}=&\sum_{\gamma=c,f} [\tilde U n_{\gamma\uparrow}n_{\gamma\downarrow}-(\mu+(-1)^\gamma t')(n_{\gamma\uparrow}+n_{\gamma\downarrow})]\nonumber\\
&+\sum_\sigma [\tilde U' n_{c\sigma}n_{f\bar\sigma}+(\tilde U'-\tilde J)n_{c\sigma}n_{f\sigma}]\nonumber\\
&-\tilde J [c^\dagger_{\downarrow}f^\dagger_{\uparrow} f_{\downarrow}c_{\uparrow}+f^\dagger_{\uparrow}f^\dagger_{\downarrow}c_{\uparrow}c_{\downarrow}+\text{h.c.}],
\label{hloc}
\end{align}
but with unconventional parameters $\tilde U=\tilde U'=\tilde J=U/2$ \cite{shinaoka2015}. In particular, the ferromagnetic Hund coupling parameter $\tilde J$ of these two-orbital systems is very large. The effect of the diagonal hopping $t'$ is to produce a chemical potential shift $\Delta\mu=\pm t'$ for the $c$ and $f$ orbitals, 
i.e., a crystal-field splitting of $\delta=2t'$ in the two-orbital model language. In  $\tilde{\mathscr{H}} _\text{loc}$, $(-1)^\gamma=1$ for the $c$ orbital and $-1$ for the $f$ orbital.

In cluster DMFT (CDMFT) \cite{lichtenstein2000}, the plaquette is coupled to a selfconsistently determined bath of noninteracting electrons, which mimics the effect of the intercluster hopping processes. The Weiss Green's function $G_0$ is the Green's function of the noninteracting embedded plaquette, and by virtue of the DMFT construction \cite{georges1996} describes the propagation via intra-cluster and inter-cluster hoppings. 
Due to the symmetries of the plaquette, 
the only nonzero elements are $(G_0)_{c_ic_j}$ ($i\ne j$) and the on-site terms $(G_0)_{c_ic_i}$ and $(G_0)_{f_if_i}$ ($i=1,2$), as shown in the third panel.

\begin{figure*}[ht]
\begin{center}
\includegraphics[angle=-90,width=0.66\columnwidth]{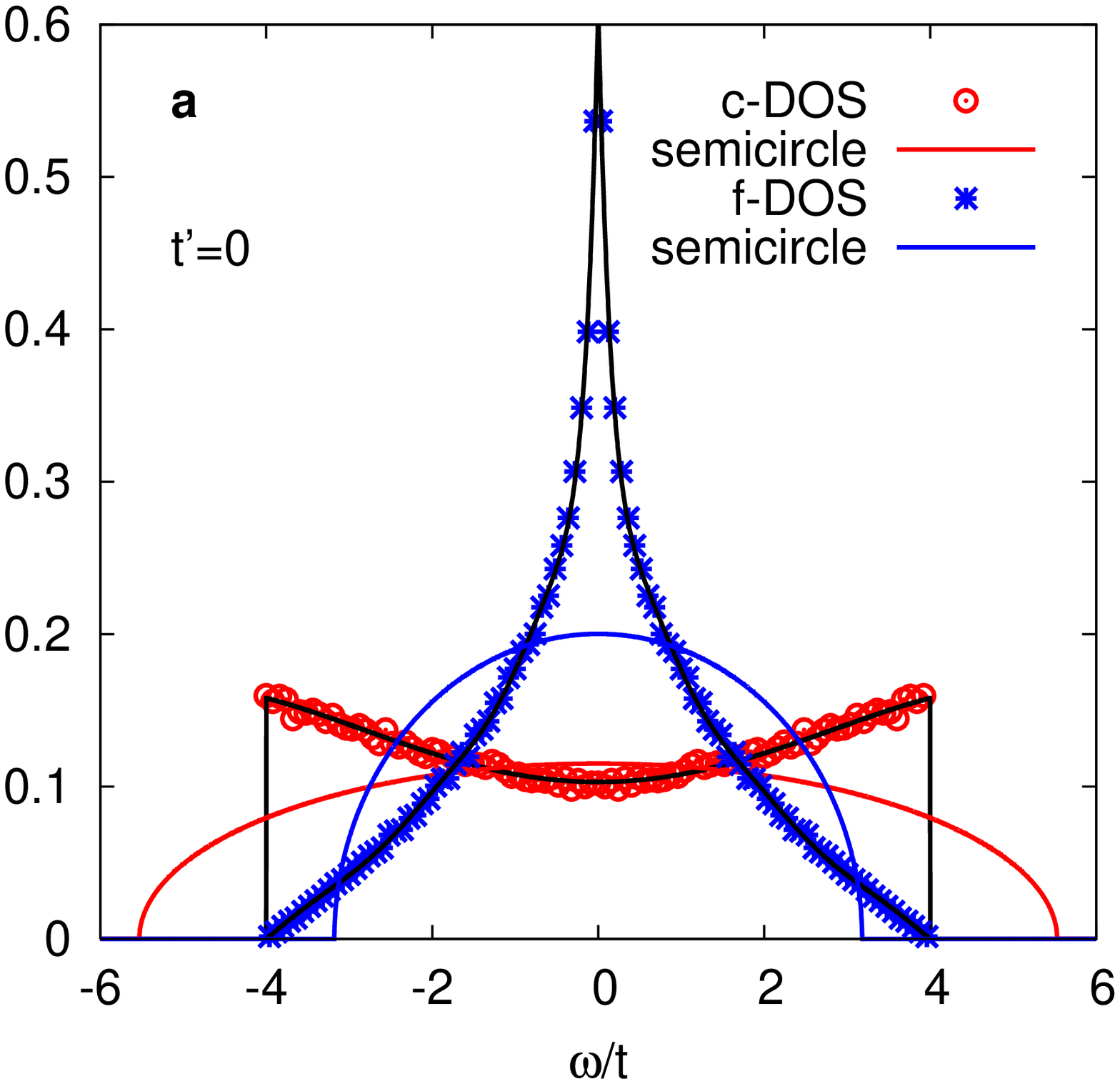}\hspace{10mm}
\includegraphics[angle=-90,width=0.66\columnwidth]{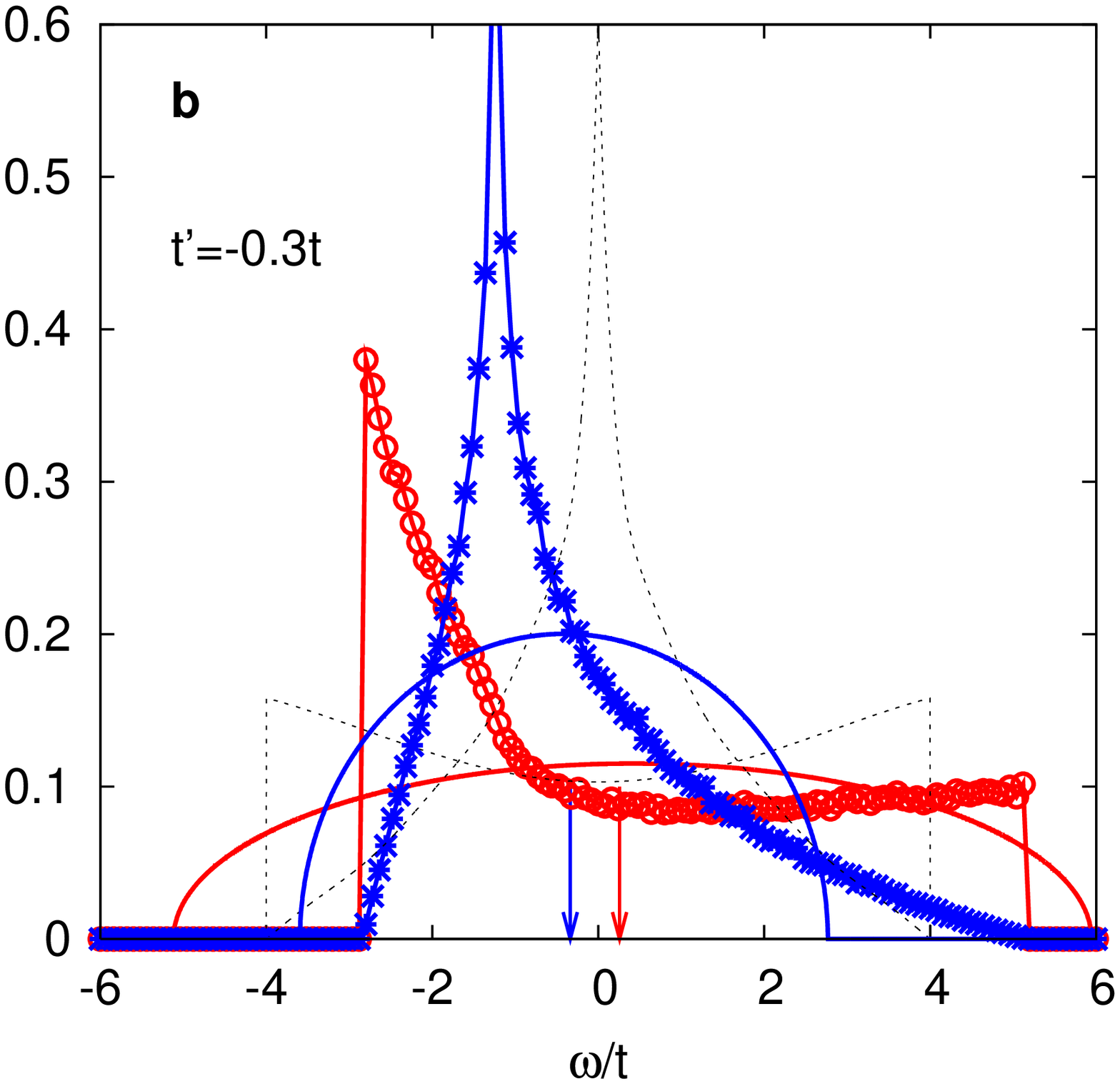}
\caption{
%(Color online) 
Noninteracting DOS for the $c$- and $f$-electrons extracted from a noninteracting plaquette CDMFT calculation (symbols) and semi-circular DOS with identical variance.
Panel {\bf a}: model with $t'=0$. Black lines show the fit functions used in the realistic-DOS simulations. 
Panel {\bf b}: Realistic DOS for the model with $t'=-0.3t$ (symbols) and semi-circular DOS with a crystal-field splitting $\delta=0.075 W$.
The arrows indicate the centers of the bands at energy $-0.34t$ and $0.26t$, respectively. 
}
\label{fig:dos}
\end{center}
\end{figure*}

The structure of $G_0$ suggests a single-site DMFT approximation based on a two-orbital model, as sketched in the right hand panel of Fig.~\ref{fig:illustration}. We thus end up with an effective description in terms of 
\begin{align}
\mathscr{H}_\text{2orbital}=&-\tilde t_c \sum_{\langle i,j\rangle,\sigma} (c^\dagger_{i,\sigma}c_{j,\sigma}+\text{h.c.})\nonumber\\
&-\tilde t_f \sum_{\langle i,j\rangle,\sigma} (f^\dagger_{i,\sigma}f_{j,\sigma}+\text{h.c.})
+\sum_i \tilde{\mathscr{H}} _{\text{loc},i},
\label{H_effective}
\end{align} 
with the local part of the Hamiltonian defined in Eq.~(\ref{hloc}) and $\tilde t_c$, $\tilde t_f$ appropriate hopping amplitudes for the $c$ and $f$ electrons. 
To obtain realistic values for the hopping parameters, we calculated the local density of states (DOS) for the $c$ and $f$ electrons from the CDMFT solution of the noninteracting model, see Fig.~\ref{fig:dos}.  If $t'=0$ (panel {\bf a}), the square-root of the variance is $2.45t$ for the $c$-DOS and $1.41t$ for the $f$-DOS. For $t'=-0.3t$ (panel {\bf b}), the $f$-DOS is shifted down, while the $c$-DOS is shifted up. 

Because single-site DMFT simulations produce the generic behavior of a high-dimensional system irrespective of the details of the DOS, we can further simplify the problem by choosing semi-circular DOS with the proper bandwidth ($W$) ratio $W_c/W_f=1.74$. 
For this choice of DOS, the DMFT selfconsistency condition becomes $\Delta_{\gamma\gamma}=(W_\gamma/4)^2 G_{\gamma\gamma}$ ($\gamma=f, c$),
where the hybridization function $\Delta$ is related to the Weiss Green's function $G_0$ by $G_{0,\gamma\gamma}^{-1}(i\omega_n)=i\omega_n+\mu-\Delta_{\gamma\gamma}(i\omega_n)$ \cite{georges1996}. We solve the DMFT equations using the matrix version \cite{werner2006matrix} of the hybridization expansion continuous-time Monte Carlo technique \cite{werner2006}. We use $W_c\equiv W$ as the unit of energy.

\vspace{2mm}\noindent
{\bf Spin-freezing and non-Fermi liquid metal}

The emergence of frozen local moments in multiorbital models with Hund coupling
profoundly affects the metal state 
close to the half-filled Mott insulator \cite{werner2008}. It is therefore interesting to explore 
the properties of model (\ref{H_effective}) as one dopes this system away from the half-filled Mott insulator. We first discuss the results obtained for the semi-circular DOS and $\delta=0$. 
To work in the relevant interaction regime of the two-orbital system, we choose $U/W=1.25$,  
which is somewhat larger than the Mott critical value $U_c/W=0.98$ $(1.03)$ of the half-filled system at inverse temperature $\beta W=200$ ($800$). 

\begin{figure*}[t]
\begin{center}
\includegraphics[angle=-90,width=0.68\columnwidth]{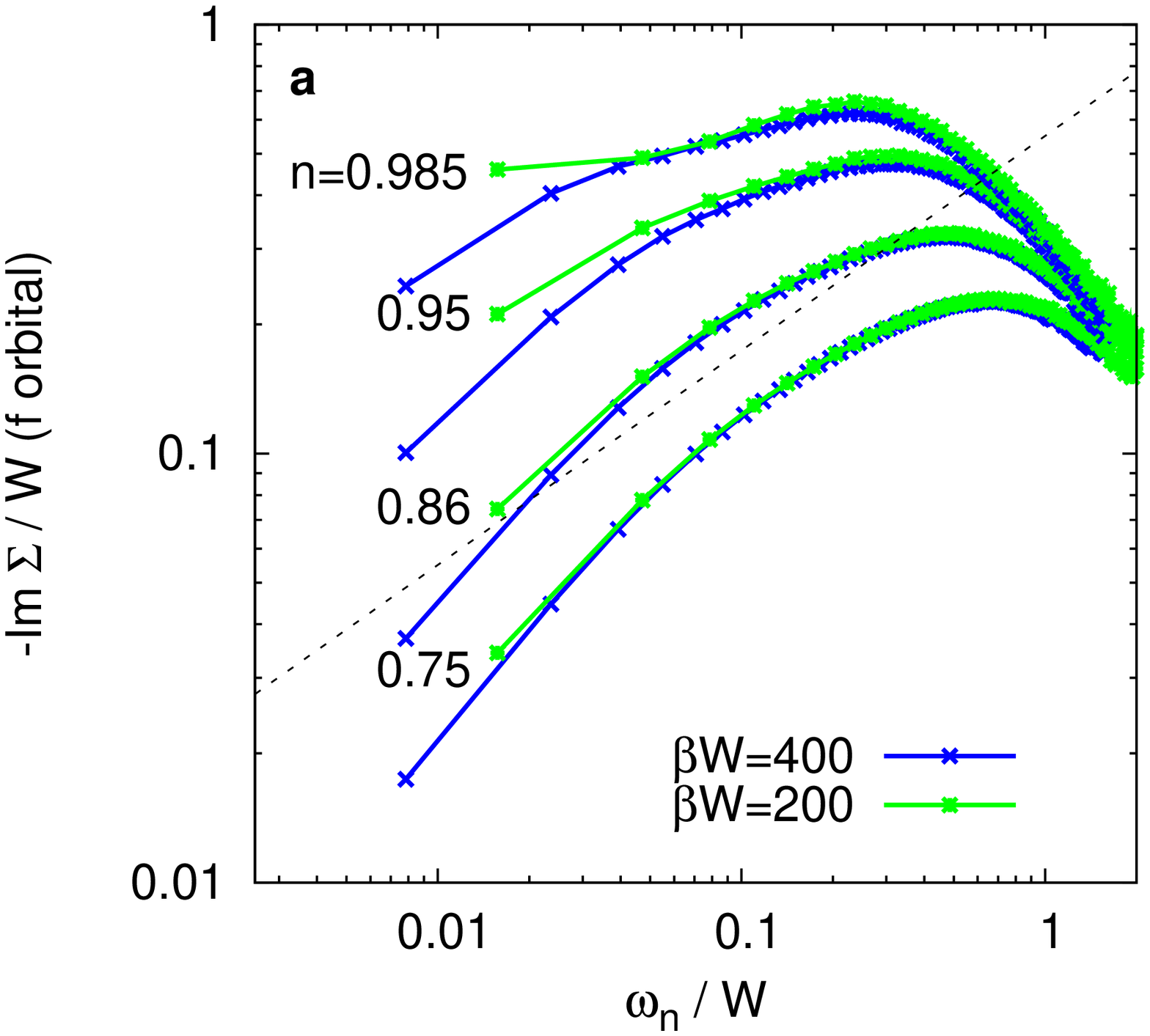}
\includegraphics[angle=-90,width=0.65\columnwidth]{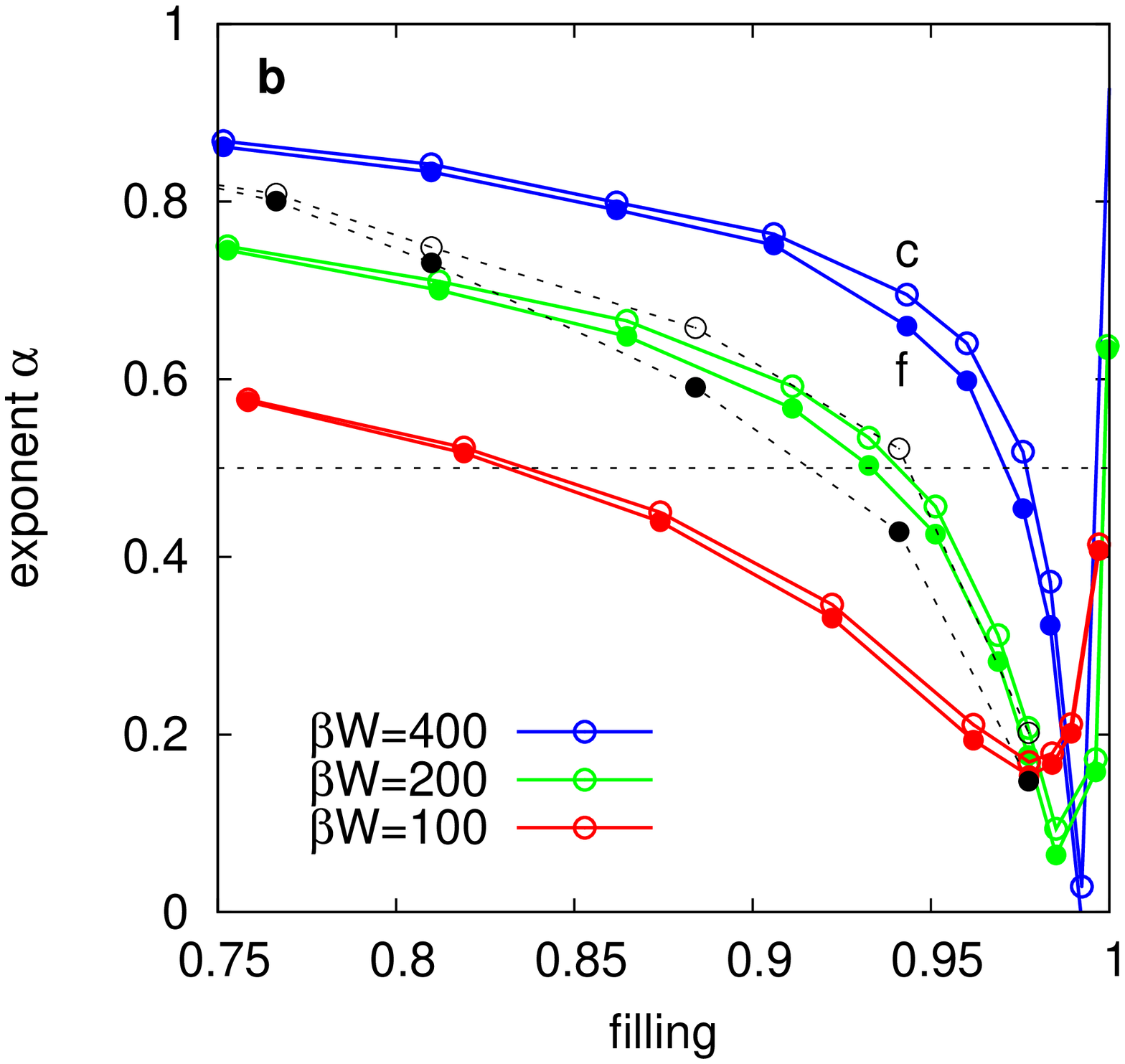}\hspace{2mm}
\includegraphics[angle=-90,width=0.65\columnwidth]{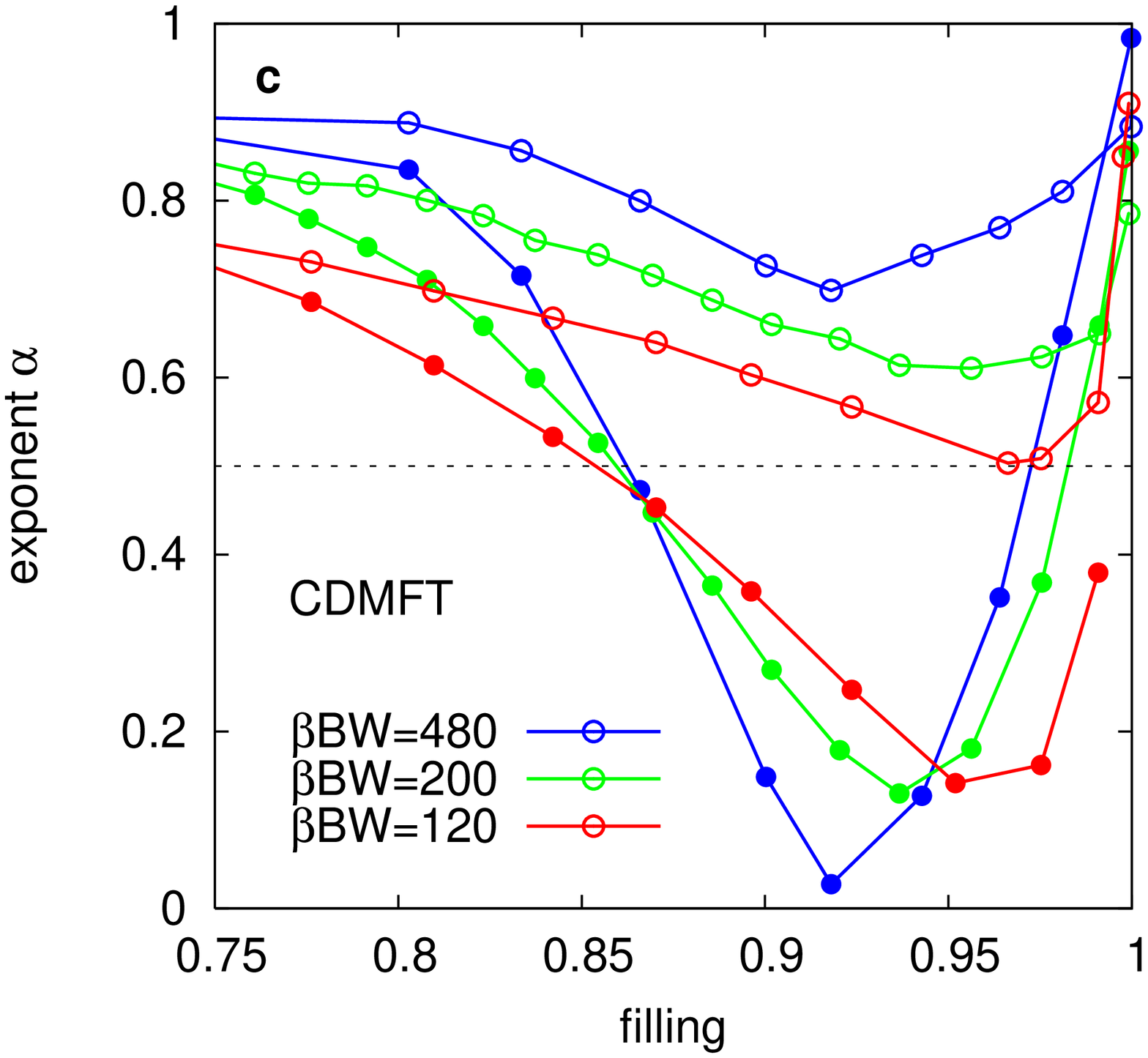}
\caption{
%(Color online) 
Doping-dependence of the self-energy revealing the non-Fermi liquid behavior and spin-freezing crossover. 
Panel {\bf a}: $f$-electron self-energies for $U/W=1.25$, $\delta=0$ and indicated inverse temperatures $\beta$. The low-frequency behavior of the self-energy indicates a crossover to a spin-frozen state around filling $n=0.98$ at the higher temperature. The black dashed line is proportional to $(\omega_n)^{1/2}$. Panel {\bf b}: Exponents $\alpha$ extracted from the fits $-\text{Im}\Sigma(i\omega_n)=b(\omega_n)^\alpha$ at the lowest two Matsubara frequencies. Both the results for the $f$ (full symbols) and $c$ electron (open symbols) are shown. For comparison, we also plot by dashed black lines the exponents obtained for the calculation with realistic DOS at $\beta W=240$. Panel {\bf c}: Analogous exponents extracted from the CDMFT calculation.  
}
\label{fig:sigma_doping}
\end{center}
\end{figure*}

Since spin-freezing physics leaves clear traces in the frequency-dependence of the self-energy \cite{werner2008} it is instructive to analyze the  
doping evolution of the 
self-energies. 
Fig.~\ref{fig:sigma_doping}{\bf a} plots $-\text{Im}\Sigma_{ff}(i\omega_n)$ for different fillings and temperatures. 
Let us characterize the low-frequency behavior by the fit $\imag \Sigma(\imu \omega_n) = b (\omega_n)^\alpha$.
The doping-dependent exponents $\alpha$, displayed in 
Fig.~\ref{fig:sigma_doping}{\bf b}, 
exhibit a minimum near half-filling, which we use to define the boundary of the spin-frozen regime. 
The minimum appears because 
the self-energy in the spin-frozen regime shows a more linear frequency dependence, similar to 
a Mott insulator with chemical potential away from the particle-hole symmetric value. 
At low temperatures, this definition of the spin-frozen regime somewhat underestimates its extent compared to the 
definition based on the fitting function $c+b (\omega_n)^\alpha$ \cite{werner2008}, 
but this detail is not important for the following discussion.

\begin{figure*}[t]
\begin{center}
\includegraphics[angle=-90,width=0.66\columnwidth]{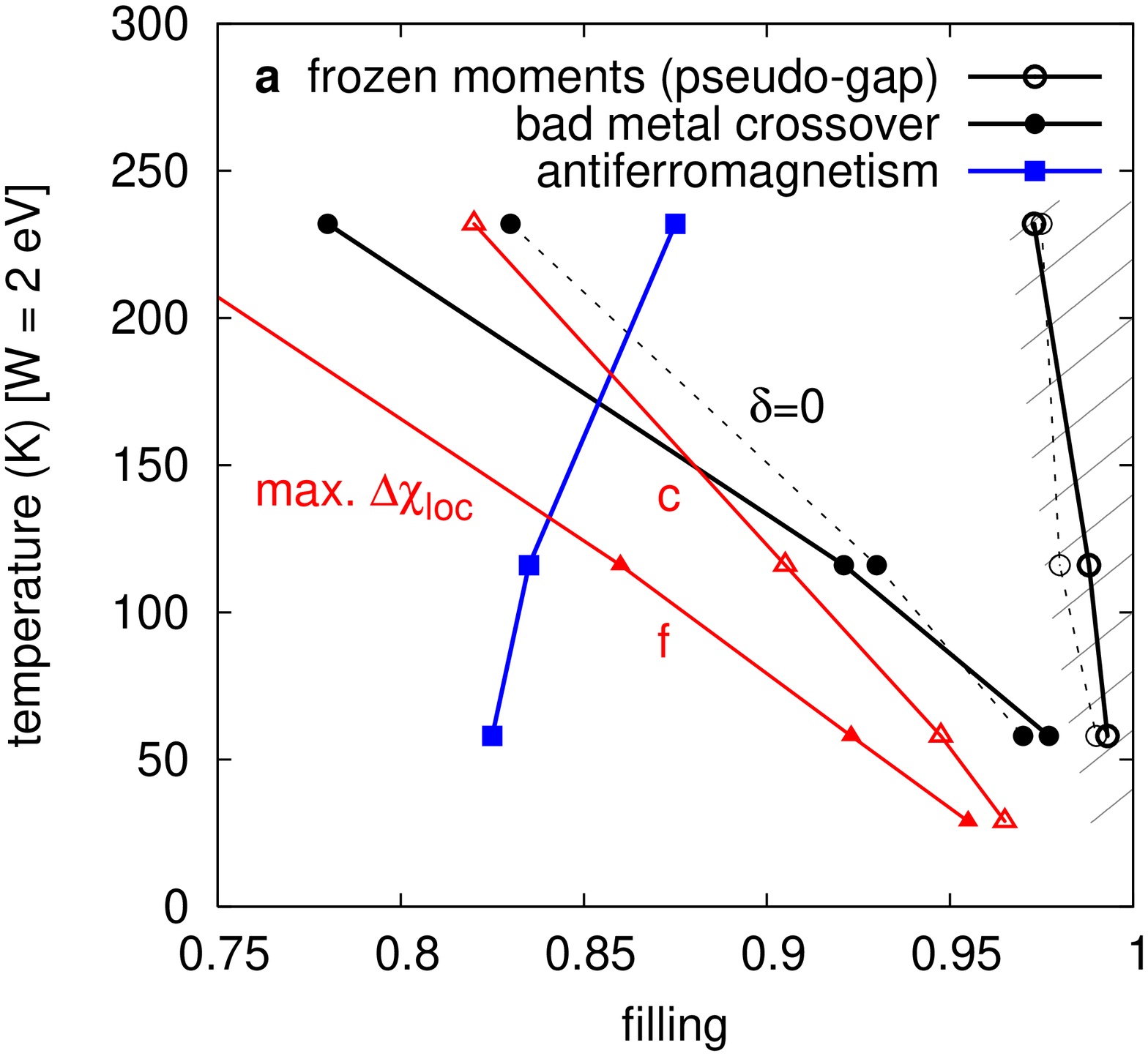}\hspace{2mm}
\includegraphics[angle=-90,width=0.66\columnwidth]{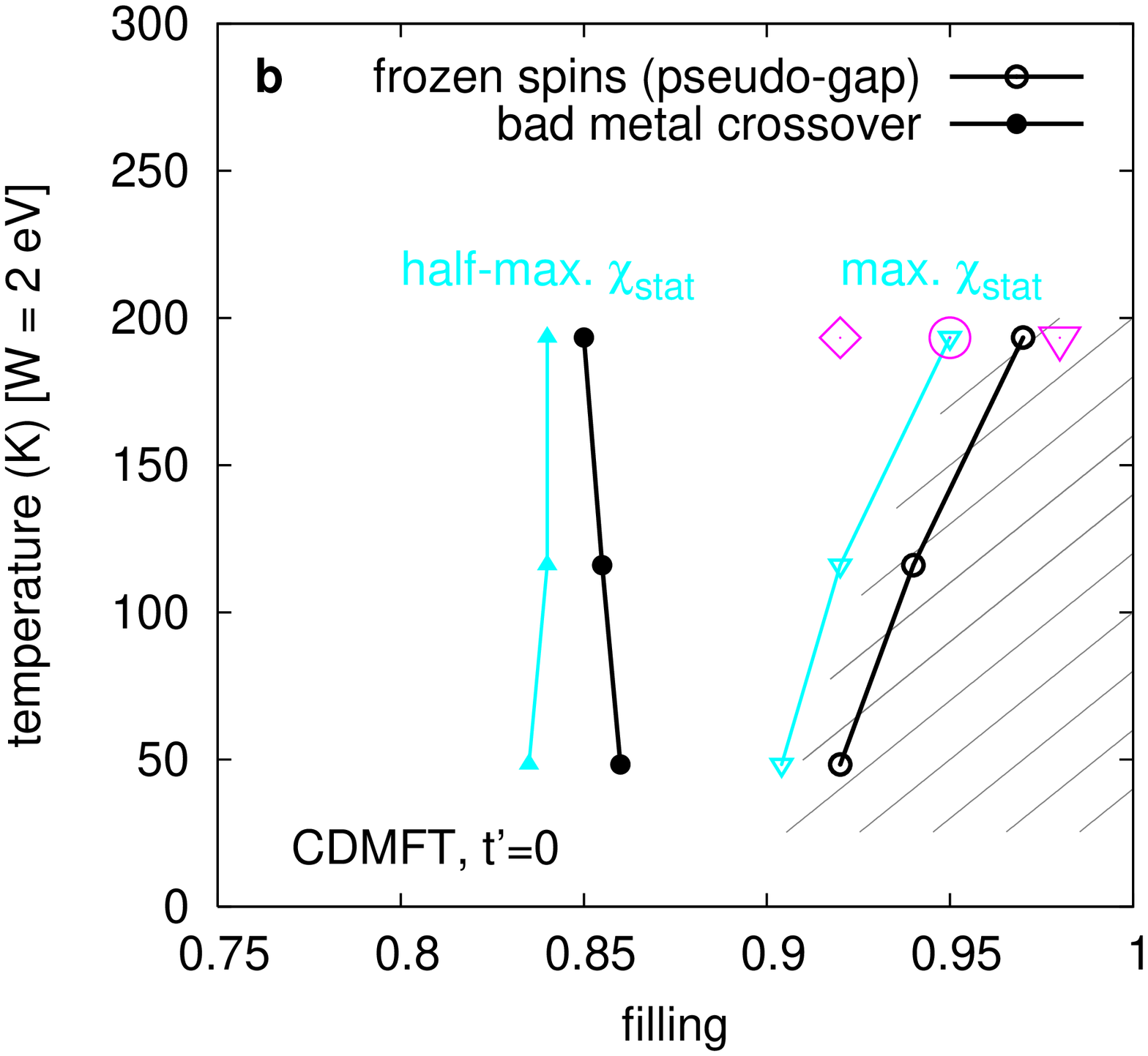}\hspace{2mm}
\includegraphics[angle=-90,width=0.66\columnwidth]{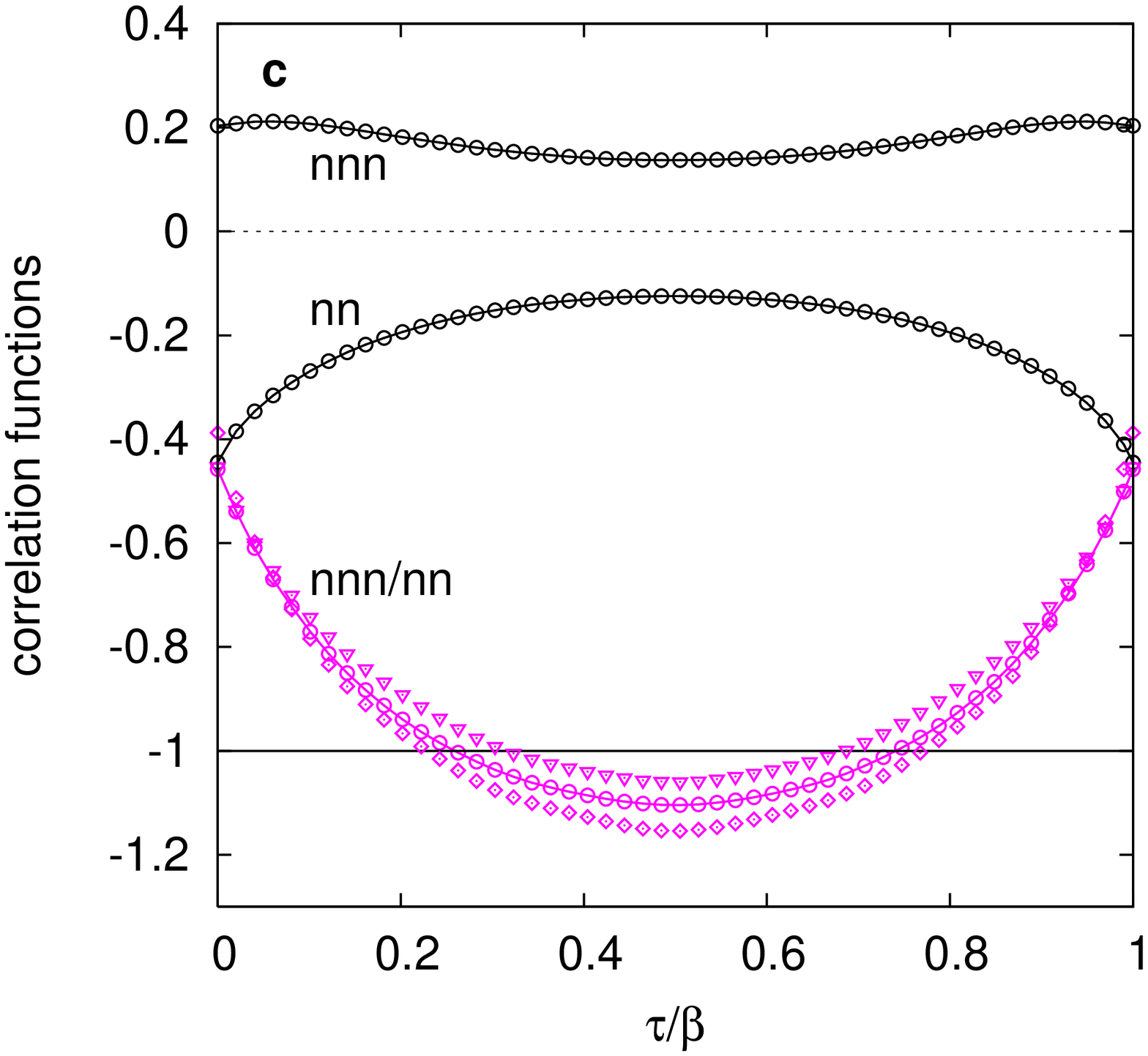}
\caption{
%(Color online) 
Phase transitions and crossover lines in the doping-temperature phase diagram. 
Here, we assume a bandwidth of 2 eV 
to translate the temperature into K. 
Panel {\bf a} shows results from single-site DMFT simulations of the effective 2-orbital model with solid lines corresponding to $\delta=0.075 W$ and dashed lines to $\delta=0$. Panel {\bf b} shows the 
crossover lines extracted from the behavior of the $f$-electron self-energy (black) and $f$-electron $S_z$-$S_z$ correlation function (light blue) in CDMFT. For three fillings near the spin-freezing line (pink symbols) we plot the ratio of next-nearest-neighbor (nnn) and nearest-neighbor (nn) $d$-electron $S_z$-$S_z$ correlations in panel {\bf c}.  
}
\label{fig:phasediagram}
\end{center}
\end{figure*}

In the semi-circle DOS calculations, the exponents extracted from the $c$- and $f$-electron self-energies are similar, with somewhat enhanced spin-freezing effects for $f$. Stable local moments exist only in a rather narrow doping range of a few percent. Within single-site DMFT, this doping range increases slightly with increasing temperature. We can also roughly determine the doping range associated with the bad metal state by using the criterion $\alpha<0.5$ for the incoherent region. 
The spin-freezing and bad-metal crossover lines are indicated by dashed black lines with open and full circles in the temperature-filling phasediagram of Fig.~\ref{fig:phasediagram}{\bf a}, where we assumed a bandwidth of 2 eV (relevant for La$_2$CuO$_4$ \cite{werner2015}), to translate temperature into K.

The single-site DMFT analysis demonstrates that our effective 2-orbital model, despite the modified Slater-Kanamori interaction with $\tilde U=\tilde U'$, the unusually large value of $\tilde J=\tilde U$ and the different bandwidths for the $c$ and $f$ electrons, exhibits the characteristic spin-freezing behavior and non-Fermi liquid 
properties expected for multi-orbital systems in the vicinity of the half-filled Mott insulator \cite{werner2008,liebsch2010,hafermann2012,hoshino2016}. Going back from the $c$/$f$- to the original $d$-fermion description, it follows that the freezing of a composite spin formed on diagonally opposite sites of the plaquette is a fundamentally important phenomenon in the 2D single-band Hubbard model (\ref{hamilt}) and, hence, in cuprates.

Let us comment on the quantitative effects of the realistic DOS. As shown in Fig.~\ref{fig:dos}, the realistic $f$-DOS has a sharp peak at $\omega=0$, 
which enhances the relative number of holes doped into the $f$-orbitals, especially at larger dopings. 
It also considerably increases the value of $U_c/W$ 
from $\approx 1$ in the semi-circle case to about $1.5$. 
Despite these quantitative changes, the spin-freezing behavior near the half-filled Mott insulator is qualitatively the same as in the semi-circle DOS simulation. To demonstrate this, we also plot the exponents $\alpha$ extracted from realistic-DOS simulations with $U=14t$ ($U=1.75W$) and $\beta t=30$ ($\beta W=240$) in Fig.~\ref{fig:sigma_doping}{\bf b} (dashed black lines). 

We will next consider the model with $t'=-0.3t$. This diagonal hopping translates into a crystal field splitting $\delta=2t'$ which pushes the $f$-band down (see Fig.~\ref{fig:dos}{\bf b}). Since $0.6t$ corresponds to $0.075$ times the bandwidth of the model with $t'=0$, we use 
such a splitting  
in the calculations with semi-circular DOS. 
The corresponding spin-freezing line is shown in Fig.~\ref{fig:phasediagram}{\bf a}  
by a solid black line with full dots, while the bad metal crossover defined by $\alpha=0.5$ is indicated by the solid black line with open dots. It turns out that the crystal field splitting does not qualitatively change the crossover lines in the temperature-filling phasediagram.

\begin{figure*}[t]
\begin{center}
\includegraphics[angle=-90,width=0.65\columnwidth]{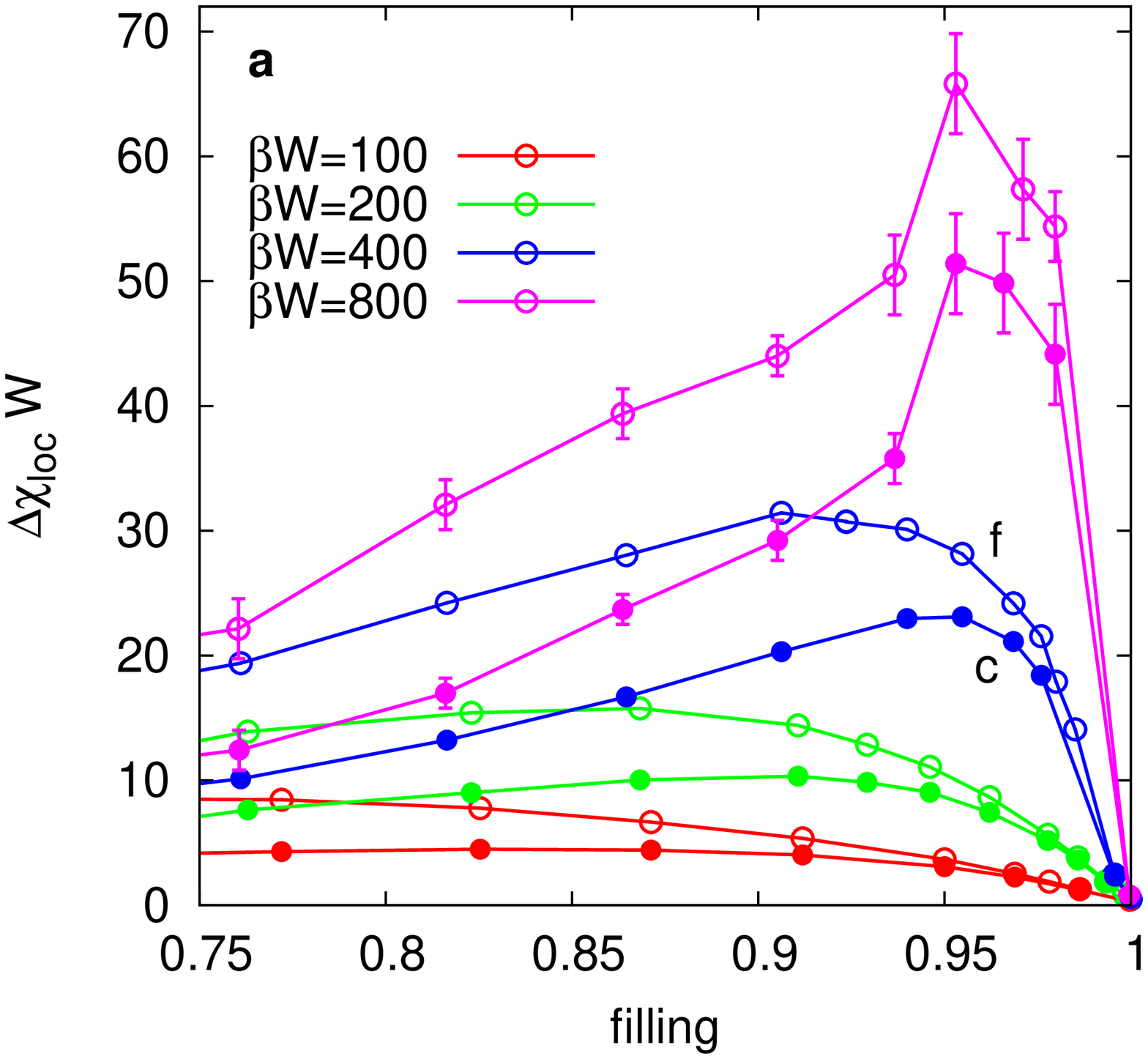}\hspace{2mm} 
\includegraphics[angle=-90,width=0.65\columnwidth]{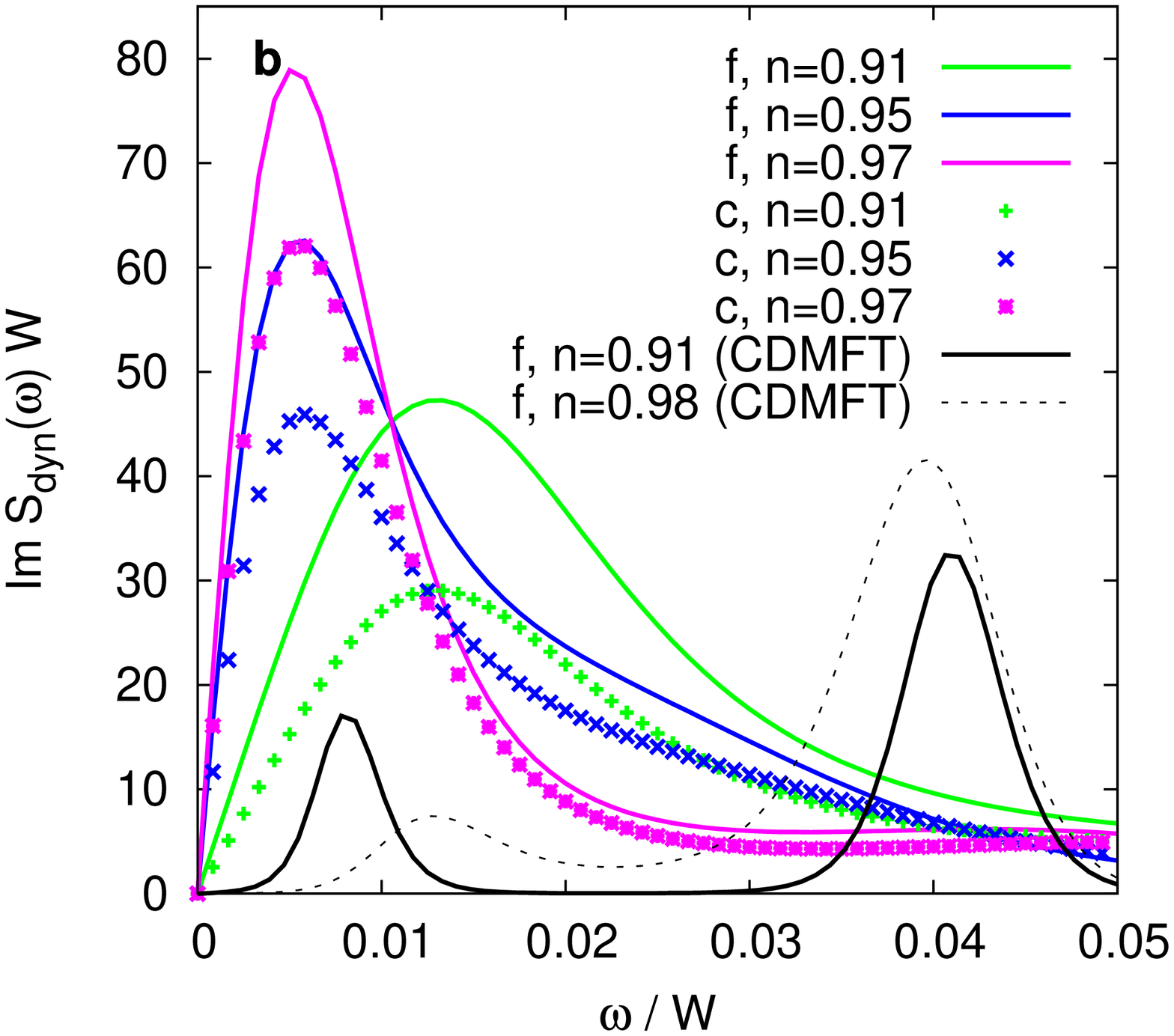}
\includegraphics[angle=-90,width=0.72\columnwidth]{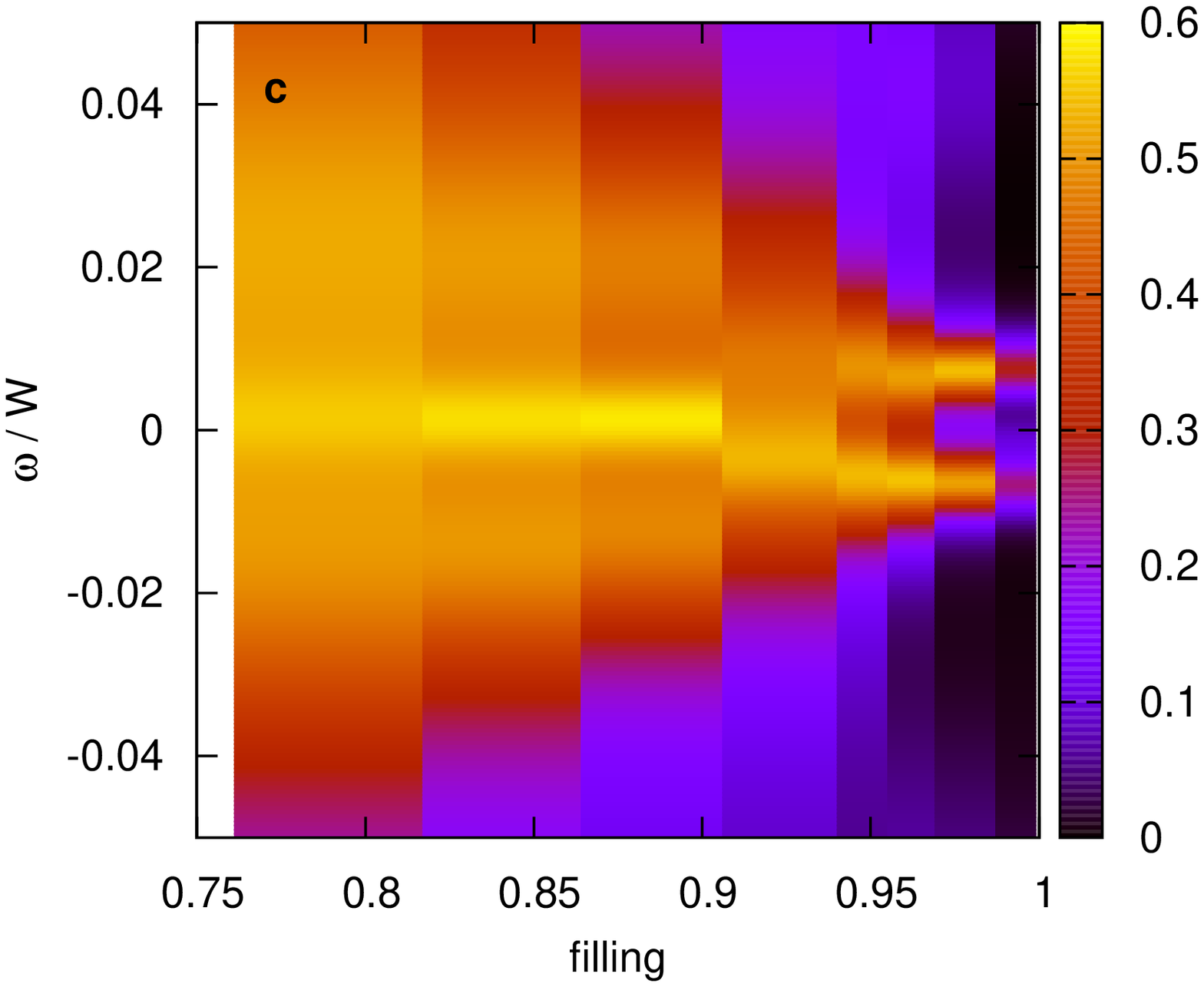}
\caption{
%(Color online) 
Local spin-susceptibility and single-particle spectral function. 
Panel {\bf a}: Dynamical contribution to 
$\chi_{\rm loc}^{(c,f)}$ 
as a function of filling for indicated values of the inverse temperature. 
Panel {\bf b}: Spectral function of the dynamical contribution to the spin-spin correlation function, 
obtained by maximum entropy analytical continuation \cite{bryan1990,lewin2016},
for $\beta W=400$ and 
different fillings. 
Panel {\bf c}: Doping evolution of the $f$-electron single-particle spectral function at $\beta W=400$. The $c$-electron spectral function, and hence also the local spectral function of the original Hubbard model show a qualitatively similar behavior. (2-orbital results for $\delta=0.075 W$, CDMFT results for $t'=0$ and $\beta W=480.$)
}
\label{fig:deltachi}
\end{center}
\end{figure*}

The bad metal behavior originates from  
Hund-coupling induced 
local moments. 
In Fig.~\ref{fig:deltachi}{\bf a} we plot the dynamical contribution to the local spin susceptibility, $\Delta \chi_{\rm loc}^{(c,f)} = \int_0^\beta d\tau S_\text{dyn}^{(c,f)}(\tau)$ with
\begin{equation}
 S_\text{dyn}^{(c,f)}(\tau)\equiv\langle S_{z}^{(c,f)}(\tau) S_{z}^{(c,f)}(0) \rangle - \langle S_{z}^{(c,f)}(\beta/2) S_{z}^{(c,f)}(0) \rangle, 
 \label{Sdyn}
 \end{equation}
 for different temperatures and dopings. The local spin fluctuations 
are strongly enhanced near the spin-frozen regime at low temperature. The peak values define the crossover lines which are plotted in red color in Fig.~\ref{fig:phasediagram}{\bf a}.
The comparison to the crossover line derived from the self-energy suggests that the non-Fermi liquid properties are caused by the slowly fluctuating local moments in the spin-freezing crossover regime. As these moments freeze below a doping concentration of 
a few percent, 
the low-energy single-particle spectral weight is strongly reduced, and a narrow pseudo-gap opens 
(see Fig.~\ref{fig:deltachi}{\bf c}). 
The size of the pseudo-gap appears to be related to the characteristic energy of the local spin fluctuations. As shown in Fig.~\ref{fig:deltachi}{\bf b}, $\text{Im}S_\text{dyn}^{(c,f)}(\omega)$ exhibits a peak near $\omega\approx 0.01 W$ in the low doping regime.

Experimentally, it is known that the normal state pseudo-gap region in cuprates can be enhanced by adding magnetic impurities  
\cite{pimenov2005}. Ellipsometry measurements showed that the addition of Ni$^{2+}$ impurities with spin $S=1$ strengthen the Cu spin correlations and induce a bulk spin-freezing transition even at optimal doping. This points to an important role of magnetic correlations in the formation of the pseudo-gap and is consistent with our spin-freezing scenario, since the static Ni$^{2+}$ moments will influence the slowly fluctuating composite spins in the spin-freezing crossover region, and (at large enough impurity concentration) lock them into a 
spin-frozen state.

\vspace{2mm}\noindent
{\bf Symmetry-breaking and short-range correlations}

It is interesting to consider also the instabilities to long-range orders and the effect of short-range correlations. Ordering instabilities can be detected by computing the corresponding lattice susceptibilities, based on a DMFT estimate of the local vertex and a solution of the Bethe-Salpeter equation \cite{hoshino2015,hoshino2016}. In the calculations with semi-circle DOS and crystal field splitting, antiferromagnetic order is stable 
at low temperature up to about 18\% 
hole doping (Fig.~\ref{fig:phasediagram}{\bf a}). As expected, the order is 
overestimated compared to CDMFT
simulations, which account for spatial fluctuations. For $U=8t$ and $t'=-0.3t$, 4-site CDMFT yields antiferromagnetic order up to 13\% 
hole doping \cite{kancharla2008}, but it was also shown that the stability region depends sensitively on 
details of the bandstructure. 

Recent studies of two-orbital models with crystal field splitting revealed an instability to spin-orbital order~\cite{kunes2015,hoshino2016}, which is intricately connected to spin-freezing.
In models with Ising type interactions 
spin-orbital order characterized by a nonzero expectation value of the operator
$o_i^{xx}=\sum_{\gamma,\gamma'=c,f}\sum_{\sigma,\sigma'} \gamma^\dagger_{i\sigma} \sigma^{x}_{\gamma\gamma'}\sigma^{x}_{\sigma\sigma'}\gamma'_{i\sigma'}$
can exist 
beyond the stability region of the antiferromagnet ($\sigma^x$ denotes a Pauli matrix) \cite{hoshino2016}.
In the presence of spin-flip and pair-hopping terms, our techniques do not allow us to search for this ordering instability. 
Nevertheless, it is interesting to note that after the transformation back to the $d$ basis, $o^{xx}$ maps onto antiferromagnetic order 
with ordering vector $\bm q=(0,\pi)$, 
represented by $\sum_{\sg\sg'}\sg^x_{\sg\sg'}(d^\dg_{1\sg}d_{1\sg'}+d^\dg_{2\sg}d_{2\sg'}-d^\dg_{3\sg}d_{3\sg'}-d^\dg_{4\sg}d_{4\sg'})$. It is degenerate with the $y,z$ components by SU(2) symmetry and with the $\bm q=(\pi,0)$ ordering vector by 90 degree rotation symmetry. 
Remarkably, {\it short-ranged} order of this type 
has been detected experimentally in cuprates upon entering the pseudo-gap phase \cite{kaminski2002,lawler2010}. 

There is also evidence from polarized neutron scattering experiments for some kind of intra-plaquette magnetic order  and time-reversal symmetry breaking in the pseudo-gap regime \cite{li2008,sidis2013}. 
While this observation has been mainly discussed in connection with the possible appearance of current loops \cite{varma2006}, there are inconsistencies between the latter theory and the experiments concerning the orientation of the moments. 
The alignment and freezing of the spins on diagonally opposite corners of the plaquette provides an alternative explanation, since it breaks time-reversal symmetry on short time- and length-scales, and reduces the 90 degree rotation symmetry to a mirror symmetry. This mechanism does not a priori favor any particular direction of the moments.

Of course, 
in the 2D Hubbard model, 
antiferromagnetic correlations and nearest-neighbor singlet formation are important and change the 
single-site DMFT results to some extent. 
To capture these effects 
one would have to implement a 2-site CDMFT calculation of the 2-orbital system (see second panel of Fig.~\ref{fig:illustration}). With a proper lattice embedding, such a simulation would be exactly equivalent to the plaquette CDMFT in the original $d$-basis. Hence, in order to address the effect of short-range correlations, we will now discuss CDMFT simulation results transformed into the $c/f$-basis, focusing on $U=8t$ and $t'=0$. 
The CDMFT simulations are performed with improved Monte Carlo updates~\cite{shinaoka2014}, 
in a single-particle basis which diagonalizes the intra-plaquette hopping. The correlation functions are measured using a worm-sampling algorithm.

As illustrated in 
Fig.~\ref{fig:sigma_doping}{\bf c},  
the nonlocal correlations result in 
a stronger differentiation between the $c$- and $f$-electron selfenergies, with the latter exhibiting much more pronounced non-Fermi liquid effects and a substantially lower ``Kondo screening" temperature. 
The second, quite expected, difference  
concerns the temperature dependence of the spin-freezing and bad-metal crossover lines. In single-site DMFT, these crossover lines have a negative slope in the temperature-filling phasediagram, because disordered local moments have a large entropy. If intersite correlations are taken into account, the frozen moments can form singlet states with a low entropy. As a result of this, the frozen moment regime (hashed region) determined from the minimum of the $f$-electron exponent $\alpha$ increases with decreasing temperature in the CDMFT solution. Similarly, the bad metal crossover line determined by the exponent $\alpha=0.5$
is now almost vertical in the temperature-filling phasediagram. 
The CDMFT crossover lines for $t'=0$ are illustrated in 
Fig.~\ref{fig:phasediagram}{\bf b}.  

The pseudo-gap regime of the CDMFT solution can still be associated with frozen $f$-moments, as evidenced by a maximum in $\chi_\text{stat}^{(f)}=\beta\langle S_z^{(f)}(\beta/2)S_z^{(f)}(0)\rangle$ near the spin-freezing crossover line,  while 
the bad metal crossover near optimal hole-doping 
is related to the emergence of local moments (the light blue line with solid triangles in Fig.~\ref{fig:phasediagram}{\bf b} indicates the doping where $\chi_\text{stat}^{(f)}$ reaches half of the maximum value). 

The main qualitative difference to the single-site 2-orbital simulations is that $\chi_\text{loc}^{(f)}$ decreases as one moves deeper into the spin-frozen regime as a result of singlet formation 
(overestimated in the 2$\times$2 geometry due to the dominance of the ``plaquette singlet state" \cite{gull2008}), 
and hence that $\Delta\chi_\text{loc}$ ceases to be a good measure of the fluctuations of the composite spin in the underdoped regime.   
To directly demonstrate the presence of robust ferromagnetic correlations along the diagonal of the plaquette we plot the nearest-neighbor and next-nearest neighbor spin correlations near the spin-frozen regime in the original $d$-basis (see Fig.~\ref{fig:phasediagram}{\bf c}). While antiferromagnetic nearest-neighbor correlations are dominant at short times, the ferromagnetic next-nearest neighbor correlations decay more slowly and eventually exceed the antiferromagnetic ones.

We have also calculated $S_\text{dyn}^{(f)}(\omega)$ from the CDMFT solution and found the same low-energy peak near $\omega\approx 0.01W$ as in the single-site 2-orbital model  (Fig.~\ref{fig:deltachi}{\bf b}, black lines). However, there is now also a second mode with an energy $\approx 0.04W$, which is related to 
antiferromagnetic fluctuations. 
In the spin-freezing crossover regime, both modes are present. In the spin-frozen (pseudo-gap) regime, the lower-energy peak associated with $f$-moment fluctuations disappears, while the peak associated with antiferromagnetic fluctuations gains weight. 

\begin{figure*}[t]
\begin{center}
\includegraphics[angle=0,width=1.5\columnwidth]{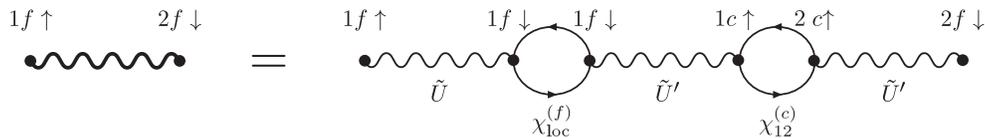}
\caption{
Illustration of the lowest order diagram which generates an attractive interaction between $(1,f,\uparrow)$ and $(2,f,\downarrow)$.
}
\label{fig:diagram}
\end{center}
\end{figure*}

\vspace{2mm}\noindent
{\bf Superconductivity}

We finally address the question of superconductivity and possible connections to spin-freezing. The 4-site CDMFT solution of the doped Hubbard model has been shown to exhibit $d$-wave superconductivity \cite{lichtenstein2000}, and we thus expect to find this ordered state in a 2-site cluster DMFT simulation of the 2-orbital model. 
Recent DMFT simulations on larger clusters 
revealed a sharp low-energy peak at $\omega\approx 0.013W$ in the imaginary part of the anomalous self-energy \cite{gull2013}. 
This energy agrees remarkably well with the characteristic energy of the local spin fluctuations, observed in both the single-site DMFT simulation of the effective 2-orbital model and in the plaquette CDMFT solution (Fig.~\ref{fig:deltachi}{\bf b}). 
This strongly suggests that the 
enhanced local moment fluctuations in the crossover regime to the spin-frozen state play a role in the formation of the $d$-wave superconducting state.

To get 
some clues of the possible  
mechanism 
let us first transform the $d$-wave order parameter $(d^\dagger_{1\uparrow}d^\dagger_{2\downarrow}-d^\dagger_{1\downarrow}d^\dagger_{2\uparrow})-(d^\dagger_{2\uparrow}d^\dagger_{3\downarrow}-d^\dagger_{2\downarrow}d^\dagger_{3\uparrow})+(d^\dagger_{3\uparrow}d^\dagger_{4\downarrow}-d^\dagger_{3\downarrow}d^\dagger_{4\uparrow})-(d^\dagger_{4\uparrow}d^\dagger_{1\downarrow}-d^\dagger_{4\downarrow}d^\dagger_{1\uparrow})$ to the $c/f$ basis. The resulting expression is remarkably simple and suggestive:
\begin{equation}
2(f^\dagger_{1\uparrow}f^\dagger_{2\downarrow}-f^\dagger_{1\downarrow}f^\dagger_{2\uparrow}),
\label{f_orderparam}
\end{equation}
where the indices $1$ and $2$ now refer to the two sites of the two-orbital cluster in the second panel of Fig.~\ref{fig:illustration}. 
It thus remains to be shown how local spin fluctuations can induce 
an effective attraction  
between the $f$-electrons with opposite spins on neighboring sites. In a weak-coupling picture \cite{inaba2012,hoshino2015} the effective interaction $\tilde U^\text{eff}_{\alpha,\beta}$ between two flavors $\alpha,\beta=(i,\gamma,\sigma)$, which takes into account simple bubble diagrams, can be obtained from the solution of the equation 
$\tilde U^\text{eff}_{\alpha\beta}=\tilde U_{\alpha\beta}-\sum_{\alpha_1} \tilde U_{\alpha\alpha_1}\chi_{\alpha_1}\tilde U^\text{eff}_{\alpha_1\beta}$. This indeed yields an attraction 
$\tilde U^\text{eff}_{(1,f,\uparrow),(2,f,\downarrow)}=2\tilde U^3\chi_\text{loc}^{(f)}\chi_{12}^{(c)}+O(\tilde U^5)$
between $f_{1,\sigma}$ and $f_{2,\bar\sigma}$ which becomes stronger with increasing $\chi_\text{loc}^{(f)}\equiv \Delta\chi_\text{loc}^{(f)}$ (in the weak-coupling regime, there are no frozen moments). 
Note that $\chi_{12}^{(c)}=-\int_0^\beta d\tau G_{c,12}(\tau)G_{c,21}(-\tau)<0$, while $\chi_\text{loc}^{(f)}>0$. 

To understand the physical mechanism, it is instructive to look at the lowest order diagram which contributes to $\tilde U^\text{eff}_{(1,f,\uparrow),(2,f,\downarrow)}$, see Fig.~\ref{fig:diagram}. 
Because of $\tilde U'=\tilde J$, the inter-orbital same-spin interaction vanishes on each site, so the $c$-$f$ interaction lines appearing in the diagram correspond to the interactions $(1,f,\sigma)$-$(1,c,\bar\sigma)$ and $(2,c,\sigma)$-$(2,f,\bar\sigma)$ (see also Fig.~\ref{fig:illustration}). Since the hopping between the sites conserves spin, the interaction between the sites is mediated by a bubble 
$\chi_{12}^{(c)}$. To connect $(1,c,\uparrow)$ to $(1,f,\uparrow)$, we have to insert a second bubble 
$\chi^{(f)}_\text{loc}$. This is how the local $f$-spin susceptibility enters the calculation, and how the enhanced local spin fluctuations increase the effective attraction between the $f$ electrons in the weak coupling approach with bubble diagrams. We note that 
the contributions to $\chi_\text{loc}$ include both the spin and charge parts, but as the interaction strength is increased, the spin contribution will dominate. 
Up to this point, our argument has only taken into account the density-density interactions. To understand why the singlet form of the order parameter  is stabilized (Eq.~(\ref{f_orderparam})), we have to consider the effect of the spin-flip 
term.

\vspace{2mm}\noindent
{\bf Conclusions}

Starting from a transformation of the 4-site Hubbard plaquette to a bonding/antibonding basis, we have derived an effective description of the 2D Hubbard model in terms of a two-orbital system with ``Slater-Kanamori" interaction and (for $t'\ne 0$) a crystal-field splitting. This model can be solved approximately within single-site DMFT, which leads to interesting new perspectives on the normal-sate properties of the Hubbard model, and hence cuprates. In particular, the two-orbital model, which features a large ferromagnetic Hund coupling, 
exhibits a spin-freezing crossover in the vicinity of the half-filled Mott insulating state. 
Our results suggest that optimally doped cuprates, like essentially all unconventional multi-band superconductors, are located in a filling and interaction regime where the normal-state properties at elevated temperature are strongly influenced by the spin-freezing phenomenon. {\it Spin/orbital-freezing thus appears to be 
a universal mechanism 
underlying the physics of (at first sight) very diverse families of unconventional superconductors, including cuprates, pnictides, ruthenates, fulleride- and uranium-based superconductors.} Specifically, for the case of cuprates, our analysis suggests that the enhanced fluctuations 
of a composite spin, consisting of aligned  
moments 
on diagonally opposite corners of a plaquette, explain the non-Fermi-liquid properties 
above the superconducting dome,  
while the freezing of these composite spins at weaker doping explains the pseudo-gap phase.  
It is interesting to note that  the  
spin-freezing scenario  
does {\it not} involve a quantum critical point since spin-freezing exists only above a certain (doping-dependent) ``Kondo screening" temperature.

Because the 
spin-frozen (pseudo-gap) state has suppressed local spin fluctuations,  
it is 
not amenable to superconductivity. In this sense, the freezing of the spins competes with superconductivity. On the other hand, the strongly and slowly fluctuating local moments 
in the spin-freezing crossover regime induce the superconducting instability and provide the glue for the $d$-wave pairing. 

As a final remark we note that spin-freezing only appears at interactions $\gtrsim U_c$, i.e., in doped Mott insulators \cite{werner2008,liebsch2010}. 
Our 
proposed mechanism thus requires that the parent compound, which in the case of the cuprates is the half-filled system, is in or close to the Mott regime.

%=================================
\vspace{2mm}\noindent
{\bf Acknowledgements}

The calculations were run on the Brutus cluster at ETH Zurich and the facilities of the Supercomputer Center at the Institute for Solid State Physics, University of Tokyo. We thank C.~Bernhard for interesting discussions, and L.~Boehnke for providing the maximum entropy code for the analytical continuation of the spin-spin correlation functions. 
HS was supported by JSPS KAKENHI Grant Numbers 16H01064 (J-Physics), 16K17735. 
PW acknowledges the hospitality of the Aspen Center for Physics and funding from FP7 ERC Starting Grant No. 278023.  
%=================================


\begin{thebibliography}{99}

\bibitem{werner2008} P. Werner, E. Gull, M. Troyer, and A. J. Millis, Phys. Rev. Lett. {\bf 101}, 166405 (2008).
\bibitem{georges2013} A. Georges, L. d. Medici, and J. Mravlje, Annual Review of Condensed Matter Physics {\bf 4}, 137 (2013).
\bibitem{hoshino2015} S. Hoshino and P. Werner, Phys. Rev. Lett. {\bf 115}, 156401 (2015).
\bibitem{hoshino2016} S. Hoshino and P. Werner, Phys. Rev. B {\bf 93}, 155161 (2016).
\bibitem{haule2009} K. Haule and G. Kotliar, New J. Phys. {\bf 11}, 025021 (2009). 
\bibitem{liebsch2010} A. Liebsch and H. Ishida, Phys. Rev. B {\bf 82}, 155106 (2010).
\bibitem{werner2012} P. Werner, M. Casula, T. Miyake, F. Aryasetiawan, A. J. Millis, and S. Biermann, Nat. Phys. {\bf 8}, 331 (2012).
\bibitem{huang2016} L. Huang and P. Werner, unpublished. 
\bibitem{capone2009} M. Capone, M. Fabrizio, C. Castellani, and E. Tosatti, Rev. Mod. Phys. {\bf 81}, 943 (2009).
\bibitem{nomura2015} Y. Nomura, S. Sakai, M. Capone, and R. Arita, Science Advances 1, e1500568 (2015).
\bibitem{steiner2016} K. Steiner, S. Hoshino, Y. Nomura, and P. Werner, arXiv:1605.06410 (2016).
\bibitem{pavarini2001} E. Pavarini, I. Dasgupta, T. Saha-Dasgupta, O. Jepsen, and O. K. Andersen, Phys. Rev. Lett. {\bf 87} 047003 (2001).
\bibitem{lichtenstein2000} A. I. Lichtenstein and M. I. Katsnelson, Phys. Rev. B {\bf 62}, 9283 (2000).
\bibitem{kotliar2001} G. Kotliar, S. Savrasov, G. Palsson, and G. Biroli, Phys. Rev. Lett. {\bf 87}, 186401 (2001).
\bibitem{maier2005} T. Maier, M. Jarrell, T. Pruschke, and M. H. Hettler, Quantum cluster theories, Rev. Mod. Phys.  {\bf 77}, 1027 (2005).
\bibitem{shinaoka2015} H. Shinaoka, Y. Nomura, S. Biermann, M. Troyer, and P. Werner, Phys. Rev. B {\bf 92}, 195126 (2015).
\bibitem{georges1996} A. Georges, G. Kotliar, W. Krauth and M. J. Rozenberg, Rev. Mod. Phys. {\bf 68}, 13 (1996).
\bibitem{werner2006matrix} P. Werner and A. J. Millis, Phys. Rev. B {\bf 74}, 155107 (2006).
\bibitem{werner2006} P. Werner, A. Comanac, L. de' Medici, M. Troyer, and A. J. Millis, Phys. Rev. Lett. {\bf 97}, 076405 (2006).
\bibitem{werner2015} P. Werner, R. Sakuma, F. Nilsson and F. Aryasetiawan, Phys. Rev. B {\bf 91}, 125142 (2015). 
\bibitem{hafermann2012} H. Hafermann, K. R. Patton, and P. Werner, Phys. Rev. B {\bf 85}, 205106 (2012).
\bibitem{pimenov2005} A. V. Pimenov, A. V. Boris, L. Yu, V. Hinkov, T. Wolf, J. L. Tallon, B. Keimer and C. Bernhard, Phys. Rev. Lett. {\bf 94}, 227003 (2005). 
\bibitem{kancharla2008} S. Kancharla et al., Phys. Rev. B {\bf 77}, 184516 (2008).
\bibitem{kunes2015} J. Kunes, J. Phys.: Condens. Matter {\bf 27}, 333201 (2015).
\bibitem{kaminski2002} A. Kaminski, S. Rosenkranz, H. M. Fretwell, J. C. Campuzano, Z. Li, H. Raffy, W. G. Cullen, H. You, C. G. Olson, C. M. Varma, and H. H\"ochst, Nature {\bf 416}, 610 (2002). 
\bibitem{lawler2010} M. J. Lawler, K. Fujita, J. Lee, A. R. Schmidt, Y. Kohsaka, C. K. Kim, H. Eisaki, S. Uchida, J. C. Davis, J. P. Sethna and E. Kim, Nature {\bf 466}, 347 (2010). 
\bibitem{li2008} Y. Li {\it et al.}, Nature {\bf 455}, 372 (2008).
\bibitem{sidis2013} S. Sidis and P. Bourges, Journal of Physics: Conference Series {\bf 449}, 012012 (2013). 
\bibitem{varma2006} C. M. Varma, Phys. Rev. B {\bf 73}, 155113 (2006).
\bibitem{shinaoka2014} H. Shinaoka, M. Dolfi, M. Troyer, and P. Werner, JSTAT P06012 (2014).
\bibitem{bryan1990} R. K. Bryan, Eur. Biophys. J. {\bf 18}, 165 (1990).
\bibitem{lewin2016} https://bitbucket.org/lewinboehnke/maxent
\bibitem{gull2008} E. Gull, P. Werner, X. Wang, M. Troyer and A. J. Millis, Europhys. Lett. {\bf 84}, 37009 (2008).
\bibitem{gull2013} E. Gull, O. Parcollet, and A. J. Millis, Phys. Rev. Lett. {\bf 110}, 216405 (2013).
\bibitem{inaba2012} K. Inaba and S. Suga, Phys. Rev. Lett. {\bf 108}, 255301 (2012).


\end{thebibliography}
\end{document}